\documentclass[english, 11pt, reqno]{amsart}
\usepackage[utf8]{inputenc}
\usepackage{hyperref, cleveref}
\usepackage{amsthm, amssymb, bm, enumitem, mathtools, mathrsfs}

\newtheorem{theorem}{Theorem}
\newtheorem{corollary}[theorem]{Corollary}
\newtheorem{lemma}[theorem]{Lemma}
\theoremstyle{remark}
\newtheorem{remark}{Remark}

\newcommand\R{\mathbb R}
\newcommand\Pro{\mathbb P}
\renewcommand{\r}{r}

\renewcommand{\geq}{\geqslant}
\renewcommand{\leq}{\leqslant}
\renewcommand{\d}{{\rm d}}
\renewcommand{\k}{{\scriptscriptstyle (k)}}
\DeclareMathOperator*{\argmin}{arg\,min}

\begin{document}

	\title[]{Asymptotics of the Kantorovich potential for the Optimal Transport with Coulomb cost}
	
	\author[R. Lelotte]{Rodrigue LELOTTE}
	\address{CEREMADE, Universit\'e Paris-Dauphine, PSL Research University, Paris, France \& SAMM, Universit\'e Paris 1 Panth\'eon-Sorbonne, Paris, France}
	\email{rodrigue.lelotte@univ-paris1.fr}
	\date{\today}

	\begin{abstract}
		We prove a conjecture regarding the asymptotic behavior 
		at infinity of the Kantorovich potential for the Multimarginal 
		Optimal Transport with Coulomb and Riesz costs. 
	\end{abstract}
	
	\maketitle
	
	\section{Introduction}

	Given a density $\rho \in L^1(\R^d, \R_+)$ with $\int_{\R^d} \rho = N$ where $N \geq 2$ is an integer, we study the following minimisation problem
	\begin{align}\label{SCE}\tag{\sffamily SCE}
		F_{\rm SCE}(\rho) := \min_{\Pro \,\mapsto \, \rho} \left\{ \int_{(\R^{d})^{N}} \sum_{1 \leq i < j \leq N} \frac{1}{|\r_i - \r_j|^s} \,\, \d \Pro(\r_1, \dots, \r_N) \right\}.
	\end{align}
	Here, $s > 0$ and the notation ``$\Pro \mapsto \rho$'' means that the minimum 
	runs over all probability measures $\Pro$ on $(\R^d)^N$ such that the marginal of $\Pro$ over the $i$-th copy of $\R^d$ in the $N$-fold Cartesian product $(\R^d)^N$ is given by $\rho/N$ for all $i = 1, \dots, N$. This is a \textit{Multimarginal Optimal Transport} (MOT) problem where all the $N$ marginals are equal and the cost of transportation is given by 
	\begin{equation}\label{eq:cost_transport}
		c(\r_1, \dots, \r_N) = \sum_{1\leq i < j \leq N} \frac{1}{|\r_i - \r_j|^s}.
	\end{equation}
	In the specific case where $d = 3$ and $s = 1$, the above cost is nothing but the \textit{Coulomb cost} in three space-dimension and the problem \eqref{SCE} is refered to as the \emph{Strictly-Correlated Electrons} functional \cite{seidl_strictly_1999, seidl_strong-interaction_1999}. Here, we will allow $s$ to take any positive scalar value, in which case the above cost of transportation is to be refered to as the \textit{Riesz cost}. By symmetry of the cost \eqref{eq:cost_transport}, we know that at least one minimiser of \eqref{SCE} is symmetric. Without loss of generality, we will therefore always restrict ourselves to symmetric $\Pro$'s hereafter.
	
	The SCE functional appears in \emph{Density-Functional Theory} (DFT), an important 
	computational method in quantum physics and chemistry. In this context, the quantity $F_{\rm SCE}(\rho)$ corresponds to the lowest possible electrostatic energy that can be reached by a system of $N$ classical electrons, under the constraint that the \textit{electronic density} of the system is given by $\rho$. We refer the reader to \cite{friesecke_strong-interaction_2023, vuckovic_density_2023} for recent surveys on the problem \eqref{SCE} and its applications in DFT.

	As a MOT problem, the problem \eqref{SCE} admits the following dual formulation, the so-called \textit{Kantorovich duality}:
	\begin{equation}\label{SCE_d_original}\tag{\sffamily KD}
		F_{\rm SCE}(\rho) = \sup_{u} \int_{\R^d} u \rho 
	\end{equation}
	where the supremum runs over all continuous functions $u : \R^d \to \R$ which are integrable with respect to $\rho$ and such that for all $\r_1, \dots, \r_N \in \R^d$ the following inequality holds 
	\begin{equation}\label{constraint_dual}
		\sum_{i = 1}^N u(\r_i) \leq \sum_{1 \leq i <j \leq N} \frac{1}{|\r_i - \r_j|^s}.
	\end{equation}
	It is proved in \cite{buttazzo_continuity_2018, colombo_continuity_2019} that a maximiser to \eqref{SCE_d_original} always exists. Maximisers of the problem \eqref{SCE_d_original} are referred to as \textit{Kantorovich potentials}. Furthermore, when the support of $\rho$ is connected -- as we will assume in our theorems -- this maximiser is actually unique on the support of $\rho$ so that one can refer to it as \textit{the} Kantorovich potential, see \cite[Prop. 4.2]{lelotteExternalDualCharge2024}.
		
	It has been argued \cite{seidl_strictly_2007, vuckovic_augmented_2017, friesecke_strong-interaction_2023} on physical ground that, when $\rho$ is supported over the entire space $\R^d$ and in the case of the 
	Coulomb potential in three space-dimension ($s = 1$ and $d = 3$), the Kantorovich potential $u$ should verify the asymptotics 
	\begin{align}\label{asymp}
		\boxed{u(\r) \overset{?}{=} \frac{N-1}{|\r|} + C_u + o\Big(\frac{1}{|r|}\Big) \quad \text{as } |\r| \to \infty.}
	\end{align}
	Here the constant $C_u$ is simply given by $C_u := \lim_{|r| \to \infty} u(r)$. This conjecture is rigorously known to hold either when $N = 2$ and $\rho$ is spherically symmetric using \cite[Cor. 3.1.1]{pass_remarks_2013} or for an arbitrary number of particles
	$N$ in one space-dimension appealing to \cite{colombo_multimarginal_2015}, using that in both cases explicit formula are available for the minimisers of \eqref{SCE}. It is also known to hold when $\rho$ 
	is compactly supported for one well-chosen Kantorovich potential, even though in this specific case, the problem is not entirely well-posed, see \cite[Rem. 3]{lelotteExternalDualCharge2024}.

	In the paper, we prove this conjecture in full generality:
	\begin{theorem}[Asymptotics of the Kantorovich potential]\label{asym_k}
		Let $\Omega \subset \R^d$ be an unbounded and connected open set, and let $\rho \in L^1(\Omega, \R_+)$ with $\int_{\R^d} \rho = N$ be such that $\rho> 0$ almost everywhere on $\Omega$. Let $u$ be the Kantorovich potential for the problem \eqref{SCE_d_original}. Then 
		\begin{align}\label{asym_th}
			u(\r) = \frac{N-1}{|\r|^s} + C_u + o\Big(\frac{1}{|r|^{s}}\Big) \quad \text{in the limit }  |\r| \to \infty\text{ and }\r \in \Omega,
		\end{align}
		for a constant $C_u$ which is negative.
	\end{theorem}
	We have thus provided a mathematically rigorous proof of the asymptotic behavior \eqref{asymp} as predicted by chemists for the Kantorovich potential. Furthermore, we have shown that this asymptotic behavior remains valid for all $s > 0$ and in any spatial dimension $d \geq 1$. In fact, our proof extends to more general interaction potentials, see Remark~\ref{rem:w} below. We note that the fact that $u$ has a well-defined limit $C_u$ at infinity was already known, see \cite[Lem. 4.1]{lelotteExternalDualCharge2024} where an explicit expression of $C_u$ is also given and where it is shown to be negative. This result is also reproduced \textit{verbatim} in Lemma~\ref{lem:C_u} below. 
	
	\begin{remark}
			In DFT,  the Kantorovich potential has been used as an approximation of the  quantum \textit{Hartree-exchange-correlation} (Hxc) potential in the context of strongly correlated materials, see \cite{malet_strong_2012, maletKohnShamDensityFunctional2013, mendl_wigner_2014, malet_density_2014} and also \cite[Sec. 4.5]{friesecke_strong-interaction_2023}. In fact, the same asymptotics as in \eqref{asymp} is predicted for the exact Hxc potential by chemists \cite[Sec. II.B]{kraislerAsymptoticBehaviorExchangeCorrelation2020} although this is a much challenging claim as this potential is not even known to exist mathematically. We note that most approximations of the Hxc potential commonly used in DFT do \textit{not} verify this asymptotics and rather yield the incorrect asymptotics $\frac{N}{|\r|}$, see \cite[Sec. V]{kraislerAsymptoticBehaviorExchangeCorrelation2020}.
	\end{remark}

			\begin{remark}\label{rem:physics_kd}
		We shall explain the physical interpretation of the Kantorovich potential. The usual optimality condition for the  Kantorovich duality in the multimarginal setting reads
	\begin{equation}\label{opt_condition}
		\sum_{1 \leq i < j \leq N} \frac{1}{|\r_i - \r_j|^s}  = \sum_{i = 1}^N u(\r_i) 
	\end{equation}
	for $\Pro$ almost-surely all $(r_1, \dots, r_N) \in (\R^d)^N$. Here, $\Pro$ is any minimiser of \eqref{SCE}. Reminiscent of \eqref{constraint_dual}, the first-order optimality condition associated to \eqref{opt_condition} reads
	\begin{equation}\label{opt_condition_grad}
		\nabla u(r_i) = \nabla_{r_i} \sum_{j \neq i } \frac{1}{|r_i - r_j|^s}
	\end{equation}
	for all $i = 1, \dots, N$ and for $\Pro$ almost-surely all $(r_1, \dots, r_N) \in (\R^d)^N$. This may be rephrased as saying that, at equilibrium, the function $u$ induces on the particle $r_i$ a force $-\nabla u (r_i)$ that exactly corresponds to that of the total repulsive force exerted onto this particle by the other $N-1$ particles of the system through the Riesz potential. Otherwise stated, the Kantorovich potential $u$ should be interpreted as the \textit{effective potential} which models the total interaction of the electrons at optimality. Another interpretation, which follows from \eqref{constraint_dual} and \eqref{opt_condition}, is that $-u$ is the \textit{external potential} which forces the particles into the density $\rho$ at optimality. 
	\end{remark}
	
	The equilibrium equation \eqref{opt_condition_grad} provides the intuition behind the asymptotic behavior of the Kantorovich potential. Indeed, if one sends an electron to infinity, it should experience the repulsive potential generated by the remaining $N - 1$ electrons -- this is precisely the meaning of the asymptotic behavior \eqref{asymp} predicted by chemists. However, the equation \eqref{opt_condition_grad} depends heavily on the support of a minimiser $\Pro$ of \eqref{SCE}, which is completely unknown. In fact, the core difficulty in proving Theorem~\ref{asym_k} lies in qualitatively understanding how the support of a minimiser behaves when a particle is pulled away from the system to infinity -- this will be the essence of Theorem~\ref{main_thm} below.
	
	In \cite{lelotteExternalDualCharge2024}, we proved that, in the Coulomb case $s = d-2$ in dimension $d > 2$, the Kantorovich potential $u$ is (up to an additive constant) the Coulomb potential induced by some positive measure $\rho_{\rm ext}$, that is
	\begin{equation}
		u(\r) = |\cdot|^{2-d} \ast \rho_{\rm ext}(\r) + C_u.
	\end{equation}
	 The measure $\rho_{\rm ext}$ is called there the \textit{dual charge}, and it is shown to be uniquely determined on the support of $\rho$ as $\rho_{\rm ext} = -c_d \Delta u$ where $$c_d := 
	 \frac{d(d-2)\pi^{d/2}}{\Gamma\left(\frac{d}{2} + 1\right)}$$ and where $\Delta$ is to be understood as the distributional Laplacian on $\R^d$. Using \Cref{asym_k}, we were able to prove the following statement.
	
	\begin{corollary}[Total mass of the dual charge]\label[corollary]{dual_charge_mass}
		We suppose that $s = d-2$ with $d > 2$. Let $\Omega \subset \R^d$ be an unbounded and connected open set, which we assume to not ``shrink in special directions at 
		infinity'', in the sense that $|A_\Omega| > 0$, where $|\cdot|$ is the Lebesgue measure on the sphere $\mathbb{S}^{d-1}$ and 
		$$
		A_\Omega := \left\{\bm{\xi} \in \mathbb{S}^{d-1} : \text{for all } \, r_0 \text{ there exists } r \geq r_0 \text{ s.t. } r\bm{\xi} \in \Omega\right\}.
		$$
		Let $\rho \in L^1(\Omega, \R_+)$ with $\int_{\R^d} \rho = N$ be such that $\rho> 0$ almost everywhere $\Omega$. Let $\rho_{\rm ext}$ be a positive measure such that 
		$|\,\cdot\,|^{2-d}\ast \rho_{\rm ext}$ is (up to an additive constant) the Kantorovich potential for the problem \eqref{SCE_d_original}. Then
		\begin{align}\label{tm}
			\int_{\R^d} \rho_{\rm ext} = N-1.
		\end{align}
	\end{corollary}
	The intuition for \Cref{dual_charge_mass} is that, an electron being fixed, the dual charge needs to exert a force which corresponds to the total repulsion due to the other $N-1$ 
	electrons, thus leading to the conjectured equality \eqref{tm}. The hypothesis regarding the shape of $\Omega$ at infinity is mainly technical and will become transparent in the proof of the corollary.

	The proofs of \Cref{asym_k} and \Cref{dual_charge_mass} are given in \Cref{proofs_as}.
	
	\vspace{1em}
	
	\emph{\textbf{Acknowledgments.}}  The author is thankful to Mathieu Lewin and Paola Gori-Giorgi, as well as to the anonymous referees for their useful comments on the first version of this work.
	
	\section{Dissociation at infinity}
	The proof of Theorem~\ref{asym_k} relies on a careful study of the support of the minimisers of \eqref{SCE} as stated in the following theorem.
	\begin{theorem}[Dissociation at infinity]\label{main_thm}
		Let $\rho \in L^1(\R^d, \R_+)$ with $\int_{\R^d} \rho = N$. Let $\Pro$ be a (symmetric) minimiser of the problem \eqref{SCE}. Then, there exists a large enough ball $B_R \subset \R^d$ of 
		radius $R$ such that 
		\begin{align}
			\label{pedantic}
			\Pro \left( (\R^d \setminus B_R) \times (\R^d \setminus B_R) \times \R^{d(N-2)}  \right) = 0.
		\end{align}
	\end{theorem}
	The above theorem (which is evidently true when $\rho$ is compactly supported) was mentioned as a conjecture in \cite[Eq. (65)]{friesecke_strong-interaction_2023}. It says that, at optimality, as one particle is sent to infinity, all the other remaining particles shall remain in a bounded domain $\Pro$ almost-surely. Otherwise stated, ``dissociation at infinity'' only occurs for one particle at a time. We will provide later in \Cref{HVZ_sec} a stronger version of this theorem.

	The proof of \Cref{main_thm} only relies on the notion of $c$-cyclical monotonicity, which we now recall. One can prove \cite{beiglbockLandMonotonePlenty2019a, zaevMongeKantorovichProblem2015} that, given any 
	minimiser $\Pro$ of \eqref{SCE}, its support $\Gamma$ is concentrated on a set $\Gamma_0$ which is \emph{$c$-cyclically monotone}, in the sense that for all $k \in \mathbb{N}$, all 
	families $(\r_1^i, \dots, \r_N^i)$ of points in $\Gamma_0$ for $i = 1, \dots, k$, and all set of permutations $\sigma_1, \dots, \sigma_N \in \mathfrak{S}_k$ we have 
	\begin{align}\label{def_cCM}
		\sum_{i = 1}^k c(\r_1^i, \dots, \r_N^{i}) \leq \sum_{i = 1}^k c(\r_1^{\sigma_1(i)}, \dots, \r_N^{\sigma_N(i)}).
	\end{align}
	Here, the meaning of \eqref{def_cCM} is that exchanging the positions of any particles at optimality necessarily leads to an increase of the energy.

	In \cite{buttazzo_continuity_2018, colombo_continuity_2019} it is proved that, at optimality, the particles cannot get too close to one another, in the sense that there exists some distance $\eta > 0$ 
	such that, for any minimiser $\Pro$ of \eqref{SCE}, we have 
	$$
	\Pro\left( \min_{1 \leq i < j \leq N} |\r_i - \r_j| < \eta \right) = 0.
	$$
	By continuity of $c$ away from the diagonals, this implies that \eqref{def_cCM} actually holds on the entire support $\Gamma$ of 
	$\Pro$. 
	Furthermore, in the formulation of the problem \eqref{SCE}, one can substitute to the Riesz potential $|\r|^{-s}$ its truncated version, \emph{i.e.} $\min \{ |\r|^{-s}, \eta^{-s}\}$. In what 
	follows, we will write $$|\r|_\eta := \max\{|\r|, \eta\},$$
	so that the truncated Riesz potential reads $|\r|_\eta^{-s}$.
	
	\begin{proof}[Proof of \Cref{main_thm}]
		We start with the two-marginal case $N=2$ as an illustration of the general argument. Let $\Pro$ be a minimiser of \eqref{SCE} with support $\Gamma$. We proceed 
		\emph{reductio ad absurdum} by assuming that, for all radii $R > 0$, we have 
		\begin{align}\label{contra_2}
			\Pro\left((\R^d \setminus B_R) \times (\R^d \setminus B_R)\right) > 0.
		\end{align}
		By definition of the support $\Gamma$ of $\Pro$, we have
		\begin{align}\label{contra_new}
			\Pro\left(\left((\R^d \setminus B_R) \times (\R^d \setminus B_R)\right) \cap \Gamma\right) > 0.
		\end{align}
		Therefore, there must exist a sequence $(\r_1^\k, \r_2^\k) \in \Gamma$ such that $\r_i^\k \to \infty$ as $k \to \infty$ ($i = 1,2$). Given any $(\r_1, \r_2) \in \Gamma$, and 
		appealing to the $c$-cyclical monotonicity, we have 
		\begin{align}\label{cCM2}
			\frac{1}{|\r_1 - \r_2|^s} \leq \frac{1}{|\r_1 - \r_2|^s} + \frac{1}{|\r_1^\k - \r_2^\k|^s} 
			\leq  \frac{1}{|\r_1 - \r_1^\k|^s} +\frac{1}{|\r_2 - \r_2^\k|^s}.
		\end{align}
		We now let $k \to \infty$ to obtain the contradiction that for all $(\r_1, \r_2) \in \Gamma$
		\begin{align}
			0 < \frac{1}{|\r_1 - \r_2|^s}  \leq 0.
		\end{align}
		Let us now consider the general case $N \geq 2$. Let $\Pro$ be any minimiser of \eqref{SCE} with support $\Gamma$. Now, let us assume that there exists a sequence of configurations in $\Gamma$ such that exactly $\mathscr{J} \in \{2, \dots, N\}$ particles escape to infinity. That is, there exists a 
		sequence $(\r_1^\k, \dots, \r_N^\k) \in \Gamma$ such that $\r_i^\k \to \infty$ as $k \to \infty$ for $i = 1, \dots, \mathscr{J}$, and such that the remaining particles remain in some 
		bounded region. Up to a subsequence and by compactness, we may assume that there exists $\overline{\r}_i \in \R^d$ such that $\r_i^\k \to \overline{\r}_i$ as $k \to \infty$ for 
		all $i = \mathscr{J} + 1, \dots, N$. Once again appealing to the $c$-cyclical monotonicity, for all $(\r_1, \dots, \r_N) \in \Gamma$ we have 
		\begin{multline}\label{cCM3}
			\sum_{1 \leq i < j \leq N} \frac1{|\r_i - \r_j|_\eta^s} + \sum_{1 \leq i < j \leq N} \frac1{|\r_i^\k - \r_j^\k|_\eta^s} \\ \leq \sum_{i = 2}^N \frac{1}{|\r_1^\k - \r_i|_\eta^s} +  \sum_{2 \leq i < 
				j \leq N} \frac1{|\r_i - \r_j|_\eta^s}  \\ + \sum_{i = 2}^N \frac{1}{|\r_1 - \r_i^\k|_\eta^s} + \sum_{2 \leq i < j \leq N} \frac1{|\r_i^\k - \r_j^\k|_\eta^s}.
		\end{multline}
		In the definition of c-cyclical monotonicity as given in \eqref{def_cCM}, the above corresponds to the choice $k = 2$ with $\sigma_1 = (12)$ and $\sigma_i = \textrm{id}$ for all 
		$i = 2, \dots, N$. Note that, contrary to \eqref{cCM2}, we used the truncated cost above. This is mainly for convenience, to emphasise that all quantities are finite. In particular, 
		after subtracting the interactions between the $\r_2, \dots, \r_N$ (resp. $\r_2^\k, \dots, \r_N^\k$) on both sides of \eqref{cCM3}, we can legally write
		\begin{align}
			\sum_{i = 2}^N \frac{1}{|\r_1 - \r_i|_\eta^s} + \sum_{i = 2}^N \frac{1}{|\r_1^\k - \r_i^\k|_\eta^s} \leq \sum_{i = 2}^N \frac{1}{|\r_1^\k - \r_i|_\eta^s} + \sum_{i = 2}^N \frac{1}{|\r_1 - 
				\r_i^\k|_\eta^s}.
		\end{align}
		We now let $k \to \infty$ to obtain that for all $(\r_1, \dots, \r_N) \in \Gamma$
		\begin{align}\label{ineq}
			\sum_{i = 2}^N \frac{1}{|\r_1 - \r_i|_\eta^s}  \leq \sum_{i = \mathscr{J}+1}^N \frac{1}{|\r_1 - \overline{\r}_i|_\eta^s}.
		\end{align}
		Now, we let $\r_i = \r_i^\k$ for $i = 1, \dots, N$ above. Letting $k \to \infty$, and by the assumption that $\mathscr{J} > 1$, we see from the above inequality that it must be that
		\begin{align}\label{asy1}
			\frac{|\r_1^\k|^s}{|\r_1^\k - \r_i^\k|^s} \to 0 \quad (i = 2, \dots, \mathscr{J}).
		\end{align}
		Indeed, to obtain the asymptotics \eqref{asy1}, let us multiply the inequality \eqref{ineq} by $|r_1|^s$. Once evaluated at $\r_i = \r_i^\k$ for $i = 1, \dots, N$, this leads to the inequality 
		\begin{equation}\label{eq:25}
			\sum_{i = 2}^\mathscr{J} \frac{|\r_1^\k|^s}{|\r_1^\k - \r_i^\k|_\eta^s} + \sum_{i=\mathscr{J}+1}^N \frac{|\r_1^\k|^s}{|\r_1^\k - \r_i^\k|_\eta^s} \leq \sum_{i = \mathscr{J}+1}^N \frac{|\r_1^\k|^s}{|\r_1 - \overline{\r}_i|_\eta^s}.
		\end{equation} 
		Now, we have the following asymptotics	\begin{equation}
			\frac{|r_1^\k|^s}{|r_1^\k - r_i^\k|^s} \to 1 \qquad \text{ and } \qquad  \frac{|r_1^\k|^s}{|r_1^\k - \overline{r}_i|^s} \to 1
		\end{equation}
		as $k\to \infty$ for the non-escaping particles, that is for all $i > \mathscr{J}$. Therefore, it follows from \eqref{eq:25} that 
		\begin{equation}
			0 \leq \limsup_{k \to \infty} \frac{|\r_1^\k|^s}{|\r_1^\k - \r_i^\k|_\eta^s} \leq 0 
		\end{equation} 
		for all $i = 2, \dots, \mathscr{J}$. This implies the aformentionned asymptotics \eqref{asy1}. But now, since
		\begin{align}\label{lo}
			\frac{|\r_1^\k|}{|\r_1^\k - \r_i^\k|} \geq \frac{|\r_1^\k|}{|\r_1^\k| + |\r_i^\k|}  =  \frac{1}{1 + \frac{|\r_i^\k|}{|\r_1^\k|}},
		\end{align}
		it follows from \eqref{asy1} that $|\r_1^\k| = o(|\r_i^\k|)$ for all $i = 2, \dots, \mathscr{J}$. But, by symmetry of $\Pro$, meaning that for any $(\r_1, \dots, \r_N) \in \Gamma$ and for any permutation $\sigma \in \mathfrak{S}_N$ we have that $(\r_{\sigma(1)}, \dots, \r_{\sigma(N)}) \in 
		\Gamma$, we can switch the indices $1$ and $i$ in \eqref{asy1}. Reproducing the above argument, we are led to $|\r_i^\k| = o(|\r_1^\k|)$ for all $i = 2, \dots, \mathscr{J}$, and therefore to a contradiction. Hence, 
		\emph{reductio ad absurdum}, the thesis of \Cref{main_thm} is proved. 
	\end{proof}
	
	\begin{remark}\label{rem:w}	
		It is easily seen from the above proof that Theorem~\ref{main_thm} remains veracious as soon as the cost of transportation is of the form
		\begin{equation}
			c(r_1, \dots, r_N) = \sum_{1 \leq i < j \leq N} w(|r_i - r_j|)
		\end{equation} 
		where the pairwise interaction potential $w : \R_+ \to (0, \infty]$ is a lower semicontinuous function that verifies, for instance, that $\lim_{ r \to \infty} w(r) = 0$ and that there exists a constant $C > 0$ and $\alpha > 0$ such that 
		\begin{equation}
			\frac{1}{Cr^\alpha} \leq w(r) \leq \frac{C}{r^\alpha} \qquad \text{ as } r \to \infty.
		\end{equation} 
	\end{remark}
	\section{Proof of the main results}\label{proofs_as}
	In this section, we will prove Theorem~\ref{asym_k} and subsequently Corollary~\ref{dual_charge_mass}. We start by briefly recalling some important facts regarding the duality theory for the multimarginal optimal transport problem \eqref{SCE}. The existence of a (continuous) maximiser for the Kantorovich duality \eqref{SCE_d_original}, a so-called Kantorovich potential, was proved originally in \cite{buttazzo_optimal-transport_2012, colombo_continuity_2019} under a \textit{small concentration} assumption on the target marginal $\rho$ -- see more precisely \cite[Assumption (A)]{buttazzo_optimal-transport_2012}. In the case where $\rho$ is in $L^1(\R^d, \R_+)$, as we assume throughout this work, this assumption is immediately verified. More precisely, the authors of \cite{buttazzo_optimal-transport_2012, colombo_continuity_2019} -- see also \cite{frieseckeOptimalTransportComprehensive2024b} -- showed that this maximiser $u$ can be chosen so as to verify the equation 
	\begin{equation}\label{eq:u}
		u(r) = \inf_{r_2, \dots, r_N \in \R^d} \left\{c(r, r_2, \dots, r_N) - \sum_{i = 2}^N u(r_i)\right\}.
	\end{equation}
	It is proved in \cite[Prop. 4.2]{lelotteExternalDualCharge2024} that the Kantorovich potential is actually unique on the connected components of the support of $\rho$. We note that it is proved there for the Coulomb cost in any dimension $d > 2$ but the extension of this result for the Riesz cost is straightforward. On this matter, we recall that two \textit{locally Lipschitz} Kantorovich potentials must agree up to additive constants on the connected components of the support of $\rho$, see \textit{e.g.} \cite[Thm. 2.14]{friesecke_strong-interaction_2023} or the proof of \cite[Prop. 4.2]{lelotteExternalDualCharge2024}. Nevertheless, this leaves out the case where there may exist a Kantorovich potential which is \textit{not} locally Lipschitz. In \cite{lelotteExternalDualCharge2024}, it is proved that this is \textit{not} possible, thus leading to the uniqueness. This entails that, in the case where the support of $\rho$ is connected, the Kantorovich potential is actually unique on this very support. Furthermore, it is proved in \cite[Lem. 4.1]{lelotteExternalDualCharge2024} that the Kantorovich potential $u$ which verifies the equation \eqref{eq:u} has a well-defined limit at infinity and this limit is identified. For the sake of convenience, we state this result here:
	\begin{lemma}[{\cite[Lemma 4.1]{lelotteExternalDualCharge2024}}]\label{lem:C_u}
		Let $u : \R^d \to \R$ be any function that verifies the equation \eqref{eq:u}. Then $u$ has a well-defined limit $C_u$ at infinity which is negative. This limit satisfies
		\begin{equation}\label{eq:C_u}
			C_u = \inf_{r_1, \dots, r_{N-1} \in \R^d} \left\{c(r_1, \dots, r_{N-1}) - \sum_{i = 1}^{N-1} u(r_i)\right\}
		\end{equation}
	\end{lemma}

	The above is proved in \cite{lelotteExternalDualCharge2024} for the Coulomb cost but this extends immediately to the case of the Riesz cost. For the sake of completeness, we provide a proof below.
	\begin{proof}[Proof of Lemma~\ref{lem:C_u}]
		If $u$ verifies the equation \eqref{eq:u}, by the non-negativity of the Riesz potential, we have that $u \geq C_u$ everywhere, where $C_u$ is, for the moment, \textit{defined} as the right-hand side of \eqref{eq:C_u}. We will now prove that $u$ has a well-defined limit at infinity and that this limit is given by $C_u$. For any $r, r_2, \dots, r_N \in \R^d$ we have by definition
		\begin{equation}\label{eq:u_smaller}
			u(r) \leq c(r, r_2, \dots, r_N) - \sum_{i = 2}^N u(r_i).
		\end{equation}
		But this implies that 
		\begin{equation}
			\limsup_{|r| \to \infty} u(r) \leq c(r_2, \dots, r_N) - \sum_{i = 2}^N u(r_i).
		\end{equation} 
		Taking the infimum with respect to $r_2, \dots, r_N \in \R^d$ in the above inequality, we obtain that $\limsup_{|r| \to \infty} u(r) \leq C_u$. Combining this with the fact that $u \geq C_u$ everywhere entails that $\lim_{|r|\to\infty} u(r) = C_u$. Finally, the fact that $C_u < 0$ is proved as follows. If we let $|r_i| \to \infty$ for all $i = 2, \dots, N$ with $|r_i - r_j| \to \infty$ for all $i \neq j \in \{2, \dots, N\}$, then it follows from \eqref{eq:u_smaller} that $C_u \leq -(N-1) C_u$, which therefore implies $C_u \leq 0$. Now, $C_u$ cannot be zero, for otherwise we would have $u \equiv 0$ --  which is evidently impossible -- according to \eqref{eq:u_smaller} and the fact that $u \geq C_u = 0$. Therefore, it must be that $C_u < 0$.
	\end{proof}
	
	\subsection{Proof of \Cref{asym_k}}
	Let $u$ be the Kantorovich potential which verifies \eqref{eq:u} and let $\Pro$ be a minimiser of \eqref{SCE} with support $\Gamma$. Let us first prove the upper
	bound asymptotics, that is
	\begin{align}\label{lba}
		u(r) \leq \frac{N-1}{|\r|^s} + C_u + o\left(\frac1{|\r|^{s}}\right).
	\end{align}
	We argue as in the proof of \cite[Thm. 1]{lelotteExternalDualCharge2024}. Let $(\r_1^\k, \dots, \r_N^\k) \in \Gamma$ be such that $\r_1^\k \to \infty$ as $k \to \infty$. Such a sequence exists as 
	the support $\Gamma$ of $\rho$ is unbounded by hypothesis. According to \Cref{main_thm} regarding the dissociation at infinity, all the other particles must stay in a bounded 
	region $\Pro$ almost-surely as $\r_1^\k$ goes to infinity. Up to a subsequence and by compactness, we may therefore assume that $\r_i^\k \to \overline{\r}_i$ for some 
	$\overline{\r}_i 
	\in \R^d$ ($i = 2, \dots, N$). We have
	\begin{align}
		\sum_{1 \leq i < j \leq N} \frac1{|\r_i ^\k- \r_j^\k|_\eta^s} - \sum_{i =1}^N u(\r_i^\k) = 0
	\end{align}
	Since $u$ is continuous and $\lim_{|r| \to \infty} u(\r) = C_u$ according to Lemma~\ref{lem:C_u}, we obtain by letting $k \to \infty$ above that
	\begin{align}\label{cs=E}
		\sum_{2 \leq i < j \leq N} \frac1{|\overline{\r}_i - \overline{\r}_j|_\eta^s} - \sum_{i = 2}^N u(\overline{\r}_i) = C_u.
	\end{align}
	We now select $\r_i = \overline{\r}_i$ ($i = 2, \dots, N$) in the equation \eqref{eq:u} verified by $u$. We obtain using the above equality that for all $\r \in \R^d$ 
	\begin{align}
		u(\r) \leq  \sum_{i = 2}^N \frac1{|\r - \overline{\r}_i|_\eta^s} + C_u
	\end{align}
	which implies the upper bound asymptotics \eqref{lba}. Now, the lower bound asymptotics is obtained as follows. By the expression \eqref{eq:C_u} of the constant $C_u$, it holds that
	\begin{align}\label{def_EN1}
		C_u \leq \sum_{2 \leq i < j \leq N} \frac1{|\r_i - \r_j|_\eta^s} - \sum_{i = 2}^N u(\r_i).
	\end{align}
	for all $ \r_2, \dots, \r_N \in \R^d$. This implies that
	\begin{equation}
		C_u \leq \sum_{1 \leq i < j \leq N} \frac1{|\r_i - \r_j|_\eta^s} - \sum_{i = 1}^N u(\r_i) - \sum_{i = 2}^N \frac{1}{|r_1 - r_i|^s_\eta} + u(r_1)
	\end{equation}
	for all $r_1 \in \R^d$ and all $\r_2, \dots, \r_N \in \R^d$. Therefore, using the optimality condition \eqref{opt_condition} we have 
	\begin{align}
		C_u \leq - \sum_{i = 2}^N \frac{1}{|r_1 - r_i|^s_\eta} + u(r_1) \quad \text{on }\, \Gamma.
	\end{align}
	By the assumption that $\rho > 0$ almost everywhere on $\Omega$, it is possible to take $|r_1| \to \infty$ with $r_1 \in \Omega$ and $(r_1, \dots, r_N) \in \Gamma$. By Theorem~\ref{main_thm}, we know that $r_2, \dots, r_N$ remain in a bounded set. Therefore, we obtain the lower bound asymptotics on $u$ as sought-after. \qed
	
	\begin{remark}\label{rem:u_not_unique}
		At the beginning of the above proof, we took $u$ to be the Kantorovich potential ``\textit{which verifies the equation \eqref{eq:u}}''. In the statement of Theorem~\ref{asym_k}, we demanded that the support of $\rho$ be connected. This implies, as already mentionned, that the Kantorovich potential $u$ is unique on the support of $\rho$ and therefore \textit{automatically} verifies the equation \eqref{eq:u} there. Nevertheless, we have kept this wording to emphasise that Theorem~\ref{asym_k} remains veracious if the support of $\rho$ is not connected. In this case the asymptotics \eqref{asym_th} is still verified for the Kantorovich potential which verifies \eqref{eq:u} -- but \textit{a priori} there may exist Kantorovich potentials which do \textit{not} verify this equation and therefore the asymptotics \eqref{asym_th}.
	\end{remark}
	
	\subsection{Proof of \Cref{dual_charge_mass}} Let us prove that if $\rho_{\rm ext}$ is a positive measure such that $U^{\rho_{\rm ext}} := |\cdot|^{d-2} \ast \rho_{\rm ext}$ is (up to an additive constant) the Kantorovich potential for the problem \eqref{SCE_d_original} where the support $\Omega$ of $\rho$ verifies the assumptions mentioned in Corollary~\ref{dual_charge_mass}, then $\rho_{\rm ext}(\R^d) = N-1$.

	The upper bound $\rho_{\rm ext}(\R^d) \leq N-1$ was already proved in \cite[Thm. 1]{lelotteExternalDualCharge2024}, 
	where it follows from the fact that, given any compactly supported (finite) measure $\mu$, we have 
	\begin{align}\label{asym_c}
		U^\mu(\r) \sim \frac{\mu(\R^d)}{|\r|^{d-2}} \quad \text{in the limit } \r \to \infty.
	\end{align}
	Nonetheless, this asymptotics need not be true for a non-compactly supported measure $\mu$\footnote{For instance, consider $\mu = \sum_{i} a_i 
		\delta_{\r_i}$ where $(\r_i)_i$ is a sequence of points such that $\r_i \to \infty$ as $i \to \infty$, and $(a_i)_i$ is a sequence of positive reals such that $\sum_{i} a_i < \infty$.}. 
	Nevertheless, one can prove \cite[Cor. 3.4]{kurokawa_order_1979} that, under the assumption
	\begin{align}\label{hyp_mu}
		\int_{\R^d} \frac{\mu(\d \r)}{|\r|^{d-2}} < \infty,
	\end{align}
	there exists a ``small'' Borel set $E \subset \mathbb{S}^{d-1}$ (in the sense that $C_d(E) = 0$, where $C_d(E)$ is the \emph{capacity} of $E$, see \cite[Chap. II]{landkof_foundations_1972}) 
	such that
	\begin{align}\label{asym_thin}
		\lim_{r \to \infty} r^{2-d} U^\mu(r \bm{\xi}) = \mu(\R^d).
	\end{align}
	for all $\bm{\xi} \in \mathbb{S}^{d-1} \setminus E$. Notice that \eqref{hyp_mu} is verified for $\rho_{\rm ext}$ as $U^{\rho_{\rm ext}}(0) < \infty$, since $U^{\rho_{\rm 
			ext}}$ is continuous. Now, by the assumption on $\Omega$ of \Cref{dual_charge_mass}, we have $|A_\Omega| > 0$ where $|\cdot|$ denotes the Lebesgue measure on the sphere $\mathbb{S}^{d-1}$ and where 
	$$
	A_\Omega := \left\{\bm{\xi} \in \mathbb{S}^{d-1} : \text{for all } \, r_0 \text{ there exists } r \geq r_0 \text{ s.t. } r\bm{\xi} \in \Omega\right\}.
	$$
	In particular, since any measurable set $B \subset \mathbb{S}^{d-1}$ of null capacity verifies $|B| = 0$ \cite[Thm. 2.1]{landkof_foundations_1972}, we have $|A_\Omega \setminus E| > 0$. 
	Therefore, there exists some direction $\bm{\xi}_0 \in A_\Omega \setminus E$ and a sequence $(r_i)_{i \geq 0}$ of positive reals such that $\lim_{i \to \infty} r_i = \infty$ and 
	\begin{align}
		\lim_{i \to \infty} r_i^{2-d} U^{\rho_{\rm ext}}(r_i \bm{\xi}_0) = \rho_{\rm ext}(\R^d), 
	\end{align}
	and such that $r_i \bm{\xi}_0 \in \Omega$ for all $i \geq 0$. Using \Cref{asym_k}, we have
	\begin{align}
		\lim_{i \to \infty} r_i^{2-d} U^{\rho_{\rm ext}}(r_i \bm{\xi}_0) = (N-1).
	\end{align}
	The thesis of \Cref{dual_charge_mass} is therefore proved. \qed
	
	\section{Dissociation and binding inequalities}\label{HVZ_sec}
	From the results established previously, we can derive several interesting consequences from a physical perspective. Given a function $v : \R^d \to \R$, let us first define 
	\begin{equation}\label{eq:def_E}
		E_K(v) := \inf_{r_1, \dots, r_K \in \R^d} \left\{\sum_{1\leq i < j \leq K} \frac{1}{|r_i - r_j|^s} + \sum_{i = 1}^K v(r_i)\right\}
	\end{equation}
	The quantity $E_K(v)$ is interpreted as the lowest possible energy that can be reached by a system of $K$ particles that interact with each other \textit{via} the Riesz potential and that are subjected to an external potential represented by the function $v$. We use the convention that $E_1(v)$ corresponds to the infimum of $v$, \textit{i.e.} $E_1(v) := \inf_{r} v(r)$, and that $E_0(v) := 0$. For latter purposes, let us also introduce the set of minimisers of \eqref{eq:def_E}, 
	\begin{equation}
		\Sigma_K(v) := \argmin_{\r_1, \dots, \r_K \in \R^d} \left\{\sum_{1\leq i < j \leq K} \frac{1}{|r_i - r_j|^s} + \sum_{i = 1}^K v(r_i)\right\}.
	\end{equation} We start by stating and proving the following lemma:
	\begin{lemma}\label{lem:hvz}
		Assume that $v : \R^d \to \R$ is continuous and that $\lim_{|r| \to \infty} v(r) = 0$. Then the energy $E_K(v)$ as defined in \eqref{eq:def_E} is non-increasing in the number of particles $K$, that is $E_K(v) \leq E_{K-1}(v)$ for all $K \geq 1$. Moreover, if the inequality is strict, that is $E_K(v) < E_{K-1}(v)$, all the minimising sequences for $E_K(v)$ are relatively compact and converge to a minimiser after extraction. In particular, the set of minimisers $\Sigma_K(v)$ is a compact subset of $(\R^d)^N$. Finally, if 
		 $$
		 v(r) = -\frac{Z}{|r|^s} + o\Big(\frac1{|r|^s}\Big), \qquad |r| \to \infty
		 $$
		for some $Z \in \mathbb{N}$ with $Z \geq 1$, then $E_K(v) < E_{K-1}(v)$ for all $1 \leq K \leq Z$.
	\end{lemma}
	 The above lemma can be thought as a sort of classical version of the celebrated \textit{HVZ theorem} due to Zhislin \cite{Zhislin-60}, Van Winter \cite{VanWinter-64} and Hunziker \cite{Hunziker-66} -- see also \cite[Thm. 6]{lewin_geometric_2011} -- which is an important and non-trivial fact for quantum systems but an easy fact for classical systems like ours. The physical intuition for the last part of Lemma~\ref{lem:hvz} is that, in the system with $K$ particles, if $Q$ particles among those are placed at infinity where the potential $v$ behaves like $-Z|\r|^{-s}$, upon those particles is exerted an attractive potential which eventually pull them back in the finite vicinity, since the remaining electrons induce a repulsive potential $(K-Q)|\r|^{-s}$ at infinity, and that $Z-(K-Q) > 0$. This is the traditional argument of Zhislin and Sigalov in \cite{Zhislin-60,zhislin_uber_1965} where the authors proved a stability result for neutral or positively-charged quantum atoms and molecules.
	
	\begin{proof}[Proof of Lemma~\ref{lem:hvz}]
		Let us first show that the energy $E_K(v)$ is non-increasing in the number of particles $K$ provided the external potential $v$ vanishes at infinity. By definition
		\begin{equation}
			E_K(v) \leq \sum_{1\leq i < j \leq K} \frac{1}{|r_i - r_j|^s} + \sum_{i = 1}^K v(r_i)
		\end{equation}
		for all $r_1, \dots, r_K \in \R^d$. Then, letting $|r_K| \to \infty$ and using that the function $v$ vanishes at infinity, we obtain that
		\begin{equation}
			E_{K}(v) \leq \sum_{1 \leq i < j \leq K - 1}  \frac{1}{|r_i - r_j|^s} + \sum_{i = 1}^{K-1} v(r_i)
		\end{equation}
		and  taking the infimum with respect to $r_1, \dots, r_{K-1}\in \R^d$ in the above equation yields the thesis. 
		
		Now, let us assume that $E_K(v) < E_{K-1}(v)$, and let $\r_1^\k, \dots, \r_K^\k$ be any minimising sequence for $E_K(v)$. Let us show that this sequence is relatively compact. Assume otherwise, so that, up to a permutation of the indices, we can extract a subsequence such that $|\r_i^\k| \to \infty$ for $i = 1, \dots, K_0$ for some $K_0 \in \{1, \dots, K\}$. Then, we have 
		\begin{align}
			E_K(v) &= \sum_{1 \leq i < j \leq K_0} \frac{1}{|r_i^\k - r_j^\k|^s} + \sum_{i = 1}^K v(r_i^\k) + o(1)\\
			&\geq \underbrace{\sum_{K_0 + 1 \leq i < j \leq K} \frac{1}{|r_i^\k - r_j^\k|^s} + \sum_{i = K_0 + 1}^K v(r_i^\k)}_{\geq E_{K - K_0}(v)} + o(1)
		\end{align}
		where we simply use the non-negavity of the Riesz potential and the fact that $v$ vanishes at infinity. Therefore, we obtain $E_{K - K_0}(v) = E_K(v) < E_{K-1}(v) \leq E_{K - K_0}(v)$ and hence a contradiction. Therefore, any minimising sequence $\r_1^\k, \dots, \r_K^\k$ for the energy $E_K(v)$ is relatively compact. By continuity, it follows that it converges to a minimiser up to extraction and that the set $\Sigma_K(v)$ is a compact subset of $(\R^d)^N$. 
		
		Finally, assume that the external potential $v$ behaves like $-Z|r|^{-s}$ at infinity from some integer $Z \geq 1$, and let us proceed by induction on the number of particles $K$ to show that $E_{K}(v) < E_{K-1}(v)$ for all $1 \leq K \leq Z$. Note that $E_1(v) < E_0(v) := 0$ since $v$ must be negative somewhere according to the asymptotics. We then proceed by induction. As an illustration of the general case, let us start by proving that $E_2(v) < E_1(v)$. First, from the fact that $E_1(v) < E_0(v)$ and that $v$ is continuous and vanishes at infinity, there exists $r^* \in \R^d$ such that $v(r^*) = E_1(v)$. By definition of $E_2(v)$, for all $\r \in \R^d$ we have
		\begin{equation}
			E_1(v) + v(\r) + \frac{1}{|\r - \r^*|^s} \geq E_2(v).
		\end{equation}
		By the asymptotic behavior of $v$, we have 
		\begin{equation}
			v(\r) + \frac{1}{|\r - \r^*|^s} = - \frac{Z - 1}{|\r|^s} + o\left(\frac{1}{|\r|^s}\right) \quad \text{as }\, |r| \to \infty.
		\end{equation}
		Therefore, for large enough $|\r| \gg 1$, we obtain $E_2(v) < E_1(v)$ and this implies according to what precedes that $E_2(v)$ is attained. The general case is treated in the same way: we assume that $E_{K-1}(v)$ is attained for some $\overline{\r}_1, \dots,\overline{\r}_{K-1}$, and we write
		\begin{equation}
			E_{K-1}(v) + v(\r) + \sum_{i = 1}^{K-1} \frac{1}{|\r - \overline{\r}_i|^s} \geq E_K(v).
		\end{equation}
		Once again, by the asymptotic behavior of $v$, we have 
		\begin{equation}
			v(\r) + \sum_{i = 1}^{K-1} \frac{1}{|\r - \overline{\r}_i|^s} = - \frac{Z - K + 1}{|\r|^s} + o\left(\frac{1}{|\r|^s}\right)\quad \text{as }\, |r| \to \infty.
		\end{equation}
		Therefore, for large enough $|\r| \gg 1$, we obtain $E_{K}(v) < E_{K-1}(v)$ -- provided that $K \geq Z$ -- and therefore that $E_K(v)$ is attained.\end{proof}
		
	We now apply the previous Lemma~\ref{lem:hvz} to the case where the external potential $v$ is chosen as $C_u - u$ where $u$ is the Kantorovich potential associated with a density $\rho$ with unbounded and connected support, and draw interesting consequences from the point of view of physics.
	\begin{corollary}[Dissociation and binding inequalities]\label{cor_HVZ}
		Let $u$ be the Kantorovich potential for the problem \eqref{SCE_d_original} for a density $\rho \in L^1(\R^d, \R_+)$ with $\int_{\R^d} \rho = N$ whose support is unbounded and connected. We consider the external potential defined by $v := C_u - u$, where we recall that $C_u := \lim_{|r|\to \infty} u(r)$. Then, we have that 
		\begin{equation}\label{diss_binding}
			E_N(v) = E_{N-1}(v) < E_{N-2}(v) < \dots < E_1(v).
		\end{equation}
		The energy $E_K(v)$ is attained for all number of particles $K = 1, \dots, N$. Furthermore, the set of minimisers $\Sigma_N(v)$ contains the support $\Gamma$ of any minimiser $\Pro$ of \eqref{SCE}. It is therefore unbounded but there exists $R > 0$ big enough such that
			\begin{equation}\label{stronger_dissociation}
				\Sigma_N(v) \cap \left( (\R^d \setminus B_R) \times (\R^d \setminus B_R) \times \R^{d(N-2)}  \right) = \emptyset.
			\end{equation}
	\end{corollary}
	
	\begin{proof}[Proof of \Cref{cor_HVZ}] This is a direct application of Lemma~\ref{lem:hvz} using that the (continuous) potential $v$ verifies $v(r) \sim - Z|r|^{-s}$ as $|r|\to\infty$ for $Z = N-1$ by Theorem~\ref{asym_k}. The item \eqref{stronger_dissociation} then follows immediately from the fact that $E_N(v) \leq E_{N-1}(v) < E_{N-K}(v)$ for all $K \geq 2$. We only need to prove that $E_N(v) = E_{N-1}(v)$ and that $E_N(v)$ is attained. However, these facts are immediate consequences of the construction. Indeed, since $u$ is a Kantorovich potential, we have that $E_N(-u) = 0$ by \eqref{constraint_dual} and \eqref{opt_condition} and this infimum is attained on the support $\Gamma$ of any minimiser $\Pro$ for the associated optimal transport problem \eqref{SCE}. This shows that $E_N(v)$ is attained. Then, since the support of $\rho$ is unbounded, we can send one (and no more than one) electron at infinity while remaining on the support of any minimiser by Theorem~\ref{main_thm}. Therefore $E_N(-u) = E_{N-1}(-u) + C_u$. This rewrites as $E_N(C_u - u) = E_{N-1}(C_u - u)$ and hence $E_N(v) = E_{N-1}(v)$. 
	\end{proof}

We conclude by commenting on the physical meaning of Corollary~\ref{cor_HVZ}. An external potential $v$ that behaves like $-(N-1)|\r|^{-s}$ at infinity can bind $N-1$ or less particles, as proved in Lemma~\ref{lem:hvz}. The fact that $E_N(v)$ is attained while $E_{N}(v) = E_{N-1}(v)$ is rather remarkable from the point of view of physics, as it implies that the potential $v$ can bind one more additional electron, leading to a total of $N$ (or less) particles. Another way to think about this fact is that one can remove an electron from the system of $N$ electrons at no cost. Furthermore, that the set of minimisers $\Sigma_N(v)$ be so big as to contain the support of any minimiser of the optimal transport problem \eqref{SCE} is also quite astonishing. Such situations are often believed to be rare and unstable, but Corollary~\ref{cor_HVZ} shows that they always occur with Kantorovich potentials.

Finally, we remark that \eqref{stronger_dissociation} is a refinement of our first dissociation theorem, namely \Cref{main_thm}. We emphasise that it need not be true that the support is equal to the whole of $\Sigma_N(v)$. For instance, in one space-dimension and for the one-dimensional Coulomb potential $-|r|$, the support can be a proper subset of $\Sigma$, see \cite[Rem. 3.5]{lelotteExternalDualCharge2024}.

\section*{Declarations}
\subsection*{Funding} This work has received funding from the 
European Research Council (ERC) under the European Union's Horizon 2020 research and innovation program (Grant agreement MDFT No 725528).
\subsection*{Competing interests}
The author has no competing interests to declare that are relevant to the content of this article.

	\bibliographystyle{acm}

\end{document}